**Comment on "Experimental Observation of Optical Rotation Generated in Vacuum by a Magnetic Field"**

In a recent Letter, it is argued [1] that a polarization rotation was observed for a light propagating in the vacuum submitted to a transverse magnetic field. Although it can be attempting to interpret this result with the existence of scalar or pseudo-scalar particles that couple to photons [1-2], even with mass and coupling constant not in the expected range for QCD axions [3] but which are claimed to be consistent with previous results [4], it seems that in the data analysis, all possible "unknown, albeit very subtle, instrumental artifacts" have not been reported or considered yet.

For example, one of the Magneto-Optical Kerr Effects (MOKE) resulting from the diamagnetic magnetization of the dielectrical mirrors of the cavity could also produce, at each reflection, a linear dichroism as well as a birefringence. In general, three different MOKE, can be distinguished as a function of the orientation of the magnetization [5] of the mirror (Fig.1). The polar one can be viewed as a Faraday effect with one reflection and in normal incidence, there is in principle no longitudinal and no transverse Kerr effects [5]. In other words, the polar MOKE is the only expected linear effect in magnetic field but from the symmetry of the PVLAS experiment, it would be induced by the smaller suspected parasitic field component. Indeed, a field component in the direction perpendicular to the mirrors is likely to arise from a slight disorientation of the superconducting magnet. Considering now the PVLAS optical rotation data [1], the ratio of the first satellite peak on the second one of their Fourier amplitude spectra is ~18 on a linear scale (25 dBV). If both peaks are due to the same polar Kerr effect, the two first harmonic stray field components perpendicular to the mirror(s) should be within the same ratio i.e. $b_{\nu_m} / b_{2\nu_m} \approx 18$. To estimate the order of magnitude of the magnetic field producing the same optical rotation as the one observed, the Verdet constants of diamagnetic materials can be used as a starting point. They are typically in the range 4-27 rad $T^{-1}$ $m^{-1}$ for $SiO_2$ type glasses and high-refraction index ones respectively [5]. With such numbers and assuming a mirror thickness of about 10 μm, a value of 8-54 $10^{-12}$ rad/mG can be obtained. Compared to the PVLAS rotation of 3.9 $10^{-12}$ rad/pass [1] and converted to the same angle but by reflection, a second harmonic magnetic field component perpendicular to each mirror in the range 0.07-0.5 mG can produce the optical rotation observed, whereas the first harmonic field component is expected to be ~18 times larger. Of course, these numbers should be precisely tuned as a function of experimental conditions and mirror characteristics. In case of a $b_{2\nu_m}$ field component much lower than the above estimate, another possibility is to consider non-linear MOKE produced by the parasitic rotating field $b_{\nu_m}$ acting either on the layers of the mirror coating or on some magnetic impurities, but both these cases are more delicate to quantify.

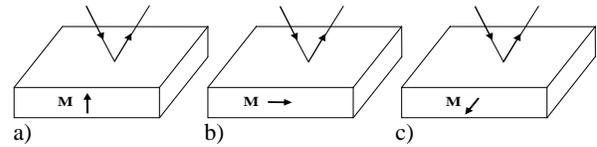

FIG 1. The three different configurations of Magneto-Optic Kerr Effects as a function of the orientation of the magnetization **M**; a) polar, b) longitudinal and c) transverse.

To rule out completely the role of MOKE in PVLAS results, precise in-situ measurements during the magnet rotation of the first and second field harmonic components in the direction parallel and perpendicular to the mirrors should be performed. It would be also important to measure the birefringence and the linear dichroism of the optical cavity with the magnetic field parallel and perpendicular to the mirrors as well as to determine the linear or quadratic field dependence.

As a summary, the answer to the role of MOKE in precision measurements probing the magnetic birefringence and linear dichroism of the vacuum with an optical cavity is of prime importance and concerns also new generation of experiments under construction [6-8]. Such a possibility was discussed during the workshop "Axions at the Institute for Advance Study" held at Princeton the 20-22 October 2006 where a similar conclusion was also presented by S. L. Adler [9].

P. Pugnat (Pierre.Pugnat@cern.ch)
CERN, CH-1211 Geneva-23, Switzerland